\documentclass[12pt]{revtex4}
\begin{document}
\title { A simple method to obtain the all order quantum corrected Bose-Einstein distribution}
\author{Anirban Bose}
\affiliation {Serampore College, Serampore, Hooghly, India.\\
}

\begin{abstract}
A simple method has been introduced to derive the all order quantum corrected Bose-Einstein distribution as the  solution of the Wigner equation. The process is a perturbative one where the Bose-Einstein distribution has been taken as the unperturbed solution. This solution has been applied to calculate the number density of the bosons at finite temperature. The study may be important to investigate the properties of bosons and bose condensates at finite temperature. This process can also be applied to  obtain the quantum corrected Fermi distribution.
\end{abstract}
\maketitle

\newpage
\section{Introduction}
In contrast to the standard approach of the wave function in the Schrodinger picture, the phase space formulation of quantum mechanics\cite{kn:neu,kn:blo,kn:wigner,kn:kirk,kn:uhl,kn:deb,kn:green} treats the position and momentum variables on equal footing and the quantum state is described by a quasiprobability diatribution function. In the existing literature, there are several approaches to find out the quasiprobability function\cite{kn:moy,kn:bon,kn:tata,kn:carr,kn:hill,kn:lee}.  Among them, perhaps the most popular distribution was discovered by Wigner\cite{kn:wigner} in 1932
in the context of many body system where he calculated the quantum correction terms to the Gibbs-Boltzmann distribution function. In later years this method has been utilized in equilibrium and non-equilibrium quantum statistical mechanics. In addition to that, it has also been applied to pure quantum mechanical problems\cite{kn:saso,kn:kors,kn:klim, kn:soto,kn:sch}. Some of them are discussed here. It has been applied to study the quantum stochastic problems\cite{kn:kors}. Calculation of  the pair distribution function of liquid neon\cite{kn:zoppi} has been performed. It has also been applied to study the quantum systems with Hamiltonian quadratic in the coordinates and momentas\cite{kn:saso,kn:klim}.Wigner distribution has turned out to be instrumental to determine the  quantum corrections to simple molecular fluids \cite{kn:singh}. It has been used to obtain the approximate solution of nonlinear Schrodinger equation\cite{kn:koro}. It has been
employed in the derivation of quantum corrections to the one component plasmas for two and three dimensional systems\cite{kn:hansen,kn:jan1}. Wigner equation is also relevant in the cotext of subjects like quantum chemistry and quantum optics\cite{kn:bas}.  

This article deals with the Wigner distribution function of indistinguishable particles. The first step in this direction was taken by Uhlhenbeck and Gropper in 1932 \cite{kn:uhl1}. They did not take into account the explicit spin effect and calculated the equation of state of nonideal Bose and Fermi gas. In the next stage, Green \cite{kn:green} worked out the connection between the density matrix obtained on the basis of classical statistics and the corresponding matrices for the bosons and fermions.  An expression for the Wigner distribution function valid for systems of bosons or
fermious is obtained by making use of correspondence relations between classical
quantities and quantum mechanical operators\cite{kn:schram}. The calculation of exchange quantum corrections in case of one component plasma has been performed \cite{kn:jan4,kn:jan3}.
Recently, there has been a renewed interest in the application of the Wigner phasespace
approach in the context of density-functional theory (DFT)\cite{kn:dahl,kn:blan}. In phase-space,
the semiclassical expansion has been applied for the inclusion of  gradient corrections to the phase-space
distribution for the spatially inhomogeneous problems to incorporate quantum corrections
beyond the local density approximation (LDA) \cite{kn:smer}.  
Another, simple closed form expression for the exact Wigner function of an ideal gas of harmonically trapped fermions or bosons at arbitrary temperature and dimensionality has been derived\cite{kn:brand}.  An introduction to
the theory along with a Monte Carlo method for the simulation of time-dependent
quantum systems of fermions  evolving in a phase-space has been presented\cite{kn:sell}.

In this work, we have calculated perturbatively  the quantum  correction to  the distribution function of the bosons. In this process we have chosen the Bose-Einstein distribution function as our unperturbed distribution. Finally, we have applied the result to calculate the number density of the bosons  by taking the proper moments of the distribution function obtained in this article. This work may be useful to deal with the properties of the particles in the Bose-Einstein condensates at finite temperature. It may be important for the construction of fluid models and determination of bulk properties of the bosons.

\section{Solution of The Complete Wigner Equation}
The Wigner equation is
\begin{eqnarray}
&&\frac{\partial f}{\partial t}+\frac{p}{m}
\frac{\partial f}{\partial {x}}-\frac{\partial
\phi}{\partial {x}}\frac{\partial f}{\partial
{p}}+ \sum_{j=1}^{\infty}(-1)^{j+1}C_{j}\hbar^{2j}\frac{\partial^{2j+1}
\phi}{\partial x^{2j+1}}
\frac{\partial^{2j+1} f}{\partial p^{2j+1}}=0 \label{v1}\end{eqnarray}

where, $f(x,p,t)$ is the single particle quasi-distribution function and $\phi$ is the potential energy.
$$C_{j}=1/{(2)}^{2j}(2j+1)!$$

The above equation will be written in a normalized form, where we have defined the following normalized variables.
$$t\sim\frac{t}{l\sqrt{m\beta}}$$
$$x\sim\frac{x•}{l•}$$ 
$$p\sim\frac{p\sqrt{\beta}}{\sqrt{m}}$$
where $l$ is the length scale of the system and $\beta$ is the Boltzmann constant.
The norlalized equation is then
\begin{eqnarray}
&&\frac{\partial f}{\partial t}+p
\frac{\partial f}{\partial {x}}-\frac{\partial
\phi}{\partial {x}}\frac{\partial f}{\partial
{p}}+ \sum_{j=1}^{\infty}(-1)^{j+1}C_{j}\Lambda^{2j}\frac{\partial^{2j+1}
\phi}{\partial x^{2j+1}}
\frac{\partial^{2j+1} f}{\partial p^{2j+1}}=0 \label{vv1}\end{eqnarray}
where $$\Lambda =\sqrt{\frac{\hbar^{2}\beta}{ml^{2}}}$$ is the small expansion parameter and this equation is  legitimate only if
it is possible to develop the potential energy $\phi$ in a Taylor series.

The semiclassical equilibrium solution will be determined in a perturbative way with $\Lambda$ as the small parameter. The starting point will be the first equation, where the first term of the infinite series of the normalized equation will be retained. The 
corresponding phase space distribution function will be denoted by $f_2$

\begin{eqnarray}
&&\frac{\partial f_{2}}{\partial t}+p
\frac{\partial f_{2}}{\partial {x}}-\frac{\partial
\phi}{\partial {x}}\frac{\partial f_{2}}{\partial
{p}}+ C_{1}\Lambda^{2}\frac{\partial^3
\phi}{\partial x^{3}}
\frac{\partial^3 f_2}{\partial p^{3}} =0 \label{v2}\end{eqnarray}

If the quantum corrections are neglected, the following expression can be obtained as the solution of the steady state vlasov equation for the indistinguishable particles

\begin{equation}f_{v}=1/[\exp(-a_{01}+a_{11}\frac{p^{2}}{2})-1]\label{v3}\end{equation}
where, $a_{01}=\phi$ and $a_{11}=1$. The chemical potential has been chosen to be zero. 

In the limit$$\exp(a_{01}-a_{11}\frac{p^{2}}{2})<1$$

$f$ can be expressed as
\begin{equation}f_{v}=\sum_{n=1}^{\infty}\exp[n(a_{01}-a_{11}\frac{p^{2}}{2})]\label{v3}\end{equation}

If the first quantum correction term is added to the vlasov equation, following solution is then obtained \cite{kn:anirban}.

\begin{equation}f_{2}=\sum_{n=1}^{\infty}W_{2n}=\sum_{n=1}^{\infty}\exp(a_{01n}-a_{11n}\frac{np^{2}}{2})\label{v3}\end{equation}
\begin{equation}a_{01n}=A+B\end{equation}
where,
$$A_{n}=-n\int\left ( 1+\frac{n^{2}{\Lambda}^{2}}{6}\frac{d^{2}\phi}{dx^{2}}\right)^{-\frac{1}{2}}\frac{d\phi}{dx}dx$$
$$B_{n}=-\frac{{n^{2}\Lambda}^{2}}{8}\int \left (1+\frac{{n^{2}\Lambda}^{2}}{6}\frac{d^{2}\phi}{dx^{2}}\right)^{-1}\frac{d^{3}\phi}{dx^{3}}dx$$
 
 and\begin{equation}a_{11n}=(1+\frac{{n^{2}\Lambda}^{2}}{6}\frac{d^{2}\phi}{dx^{2}})^{-\frac{1}{2}}\label{v5}\end{equation}
 
 If, terms upto $\Lambda^{2}$ order are retained
\begin{equation}a_{01n}=- n\phi +\frac{n^{3}\Lambda^{2}}{24}\frac{d\phi}{dx}\frac{d\phi}{dx}-\frac{n^{2}\Lambda^{2}}{8}\frac{d^{2}\phi}{dx^{2}}\label{4}\end{equation}

\begin{equation}a_{11n}=
1-\frac{ n^{2}\Lambda^{2}}{12}\frac{d^{2}\phi}{dx^{2}}\label{5}\end{equation}

In the next stage, the first two terms of the series will be taken into account and a solution will be sought of the following form

$$f_{4}=\sum_{n=1}^{\infty}W_{2n}W_{4n}$$

where,

\begin{equation}W_{4n}=\exp{[\Lambda^{4}(a_{02n}+a_{12n}\frac{np^{2}}{2}+a_{22n}{{(\frac{np^{2}}{2})}^{2}})]}\label{v6}\end{equation}

The postulated solution will be inserted in the following equation
\begin{eqnarray}
&&\frac{\partial f_{4}}{\partial t}+p
\frac{\partial f_{4}}{\partial {x}}-\frac{\partial
\phi}{\partial {x}}\frac{\partial f_{4}}{\partial
{p}}+ C_{1}\Lambda^{2}\frac{\partial^3
\phi}{\partial x^{3}}
\frac{\partial^3 f_4}{\partial p^{3}}-C_{2}\Lambda^{4}\frac{\partial^5
\phi}{\partial x^{5}}
\frac{\partial^5 f_4}{\partial p^{5}} =0 \label{v7}\end{eqnarray}

The method can be illustrated with the fourth term of the above equation. It can be written explicitly.
$$C_{1}\Lambda^{2}\frac{\partial^3
\phi}{\partial x^{3}}
\frac{\partial^3 f_4}{\partial p^{3}}=\sum_{n=1}^{\infty} C_{1}\Lambda^{2}\frac{\partial^3
\phi}{\partial x^{3}}
\frac{\partial^3 W_{2n}}{\partial p^{3}}W_{4n} +\sum_{n=1}^{\infty}3C_{1}\Lambda^{2}\frac{\partial^3
\phi}{\partial x^{3}}
\frac{\partial^2 W_{2n}}{\partial p^{2}}\frac{\partial W_{4n}}{\partial p}+$$
$$\sum_{n=1}^{\infty}3C_{1}\Lambda^{2}\frac{\partial^3
\phi}{\partial x^{3}}
\frac{\partial W_{2n}}{\partial p}\frac{\partial^{2} W_{4n}}{\partial p^{2}}+\sum_{n=1}^{\infty}3C_{1}\Lambda^{2}\frac{\partial^3
\phi}{\partial x^{3}}
 W_{2n}\frac{\partial^{3} W_{4n}}{\partial p^{3}}$$
Similarly,
$$C_{2}\Lambda^{4}\frac{\partial^5
\phi}{\partial x^{5}}
\frac{\partial^5 f_4}{\partial p^{5}}=\sum_{n=1}^{\infty}C_{2}\Lambda^{4}\frac{\partial^5
\phi}{\partial x^{5}}
\frac{\partial^5 W_{2n}}{\partial p^{5}}W_{4n} +...+\sum_{n=1}^{\infty}C_{2}\Lambda^{4}\frac{\partial^5
\phi}{\partial x^{5}}
 W_{2n}\frac{\partial^{5} W_{4n}}{\partial p^{5}}$$
It can be noticed from the expression of the $W_{4n}$ that its derivative will bring out $\Lambda^{4}$ from the argument of the exponential and will be atleast of the order of $\Lambda^{4}$. Therefore, except the first terms of the above expressions all the other terms are  of higher  order than  $\Lambda^{4}$ and will be omitted.
If the terms upto the order $\Lambda^{4}$ are retained, then 
\begin{eqnarray}
&&\sum_{n=1}^{\infty}(p
\frac{\partial W_{2n}}{\partial {x}}-\frac{\partial
\phi}{\partial {x}}\frac{\partial W_{2n}}{\partial
{p}}+ C_{1}\Lambda^{2}\frac{\partial^3
\phi}{\partial x^{3}}
\frac{\partial^3 W_{2n}}{\partial p^{3}})W_{4n}-\sum_{n=1}^{\infty} C_{2}\Lambda^{4}\frac{\partial^5
\phi}{\partial x^{5}}\frac{\partial^5 W_{2n}}{\partial p^{5}}W_{4n}\nonumber \\
&+&\sum_{n=1}^{\infty}(p\frac{\partial W_{4n}}{\partial {x}}-\frac{\partial
\phi}{\partial {x}}\frac{\partial W_{4n}}{\partial
{p}})W_{2n} =0 \label{v8}\end{eqnarray}
Finally, eq.
If the value of $W_{2n}$ corrected upto $\Lambda^{2}$ is inserted in this equation, higher order terms including $\Lambda^{2}$ order will be generated from the third term. The terms upto the $\Lambda^{2}$ order will be omitted from the first three terms and the following two $\Lambda^{4}$ order term will contribute to the $\Lambda^{4}$ order solution.  
$$-\Lambda^{4}C_{1}\sum_{n=1}^{\infty}\frac{n^{4}•}{2•}\frac{d^{2}\phi}{dx^{2}•}\frac{d^{3}\phi}{dx^{3}•}pW_{2n}W_{4n}$$
$$\Lambda^{4}C_{1}\sum_{n=1}^{\infty}\frac{n^{5}•}{4•}\frac{d^{2}\phi}{dx^{2}•}\frac{d^{3}\phi}{dx^{3}•}p^{3}W_{2n}W_{4n}$$
Now, eq.(\ref{v8}) will be 
\begin{eqnarray}
&&-\Lambda^{4}C_{1}\sum_{n=1}^{\infty}\frac{n^{4}•}{2•}\frac{d^{2}\phi}{dx^{2}•}\frac{d^{3}\phi}{dx^{3}•}pW_{2n}W_{4n}+\Lambda^{4}C_{1}\sum_{n=1}^{\infty}\frac{n^{5}•}{4•}\frac{d^{2}\phi}{dx^{2}•}\frac{d^{3}\phi}{dx^{3}•}p^{3}W_{2n}W_{4n}
-\sum_{n=1}^{\infty} C_{2}\Lambda^{4}\frac{\partial^5
\phi}{\partial x^{5}}\frac{\partial^5 W_{2n}}{\partial p^{5}}W_{4n}\nonumber \\
&+&\sum_{n=1}^{\infty}(p\frac{\partial W_{4n}}{\partial {x}}-\frac{\partial
\phi}{\partial {x}}\frac{\partial W_{4n}}{\partial
{p}})W_{2n} =0 \label{v81}\end{eqnarray}
Inserting $W_{2n}$ and $W_{4n}$ from eq.(\ref{v3}) and eq.(\ref{v6}) respectively and collecting the $\Lambda^{4}$ order coefficients of different powers of p and separately equating them to zero, the following first order differential equations emerge

$p^{5}\longrightarrow$
\begin{equation}\frac{\partial a_{22n}}{\partial x}=-4n^{3}C_{2}\frac{\partial^5
\phi}{\partial x^{5}}\label{v9}\end{equation}

$p^{3}\longrightarrow$
\begin{equation}\frac{\partial a_{12n}}{\partial x}=20n^{3}C_{2}\frac{\partial^5
\phi}{\partial x^{5}}+2n\frac{\partial
\phi}{\partial {x}}a_{22n}-C_{1}\frac{n^{4}}{2}\frac{\partial^{2}
\phi}{\partial {x^{2}}}\frac{\partial^{3}
\phi}{\partial {x^{3}}}\label{v10}\end{equation}

$p\longrightarrow$
\begin{equation}\frac{\partial a_{02n}}{\partial x}=n\frac{\partial
\phi}{\partial x}a_{12n}-15n^{3}C_{2}\frac{\partial^{5}
\phi^{}}{\partial {x}^{5}}+C_{1}\frac{n^{4}}{2}\frac{\partial^{2}
\phi}{\partial {x^{2}}}\frac{\partial^{3}
\phi}{\partial {x^{3}}}\label{v11}\end{equation}

From  eq.(\ref{v9})
\begin{equation}a_{22n}=-4n^{3}C_{2}\frac{\partial^4
\phi}{\partial x^{4}}\label{v12}\end{equation}

\begin{equation}a_{12n}=20n^{3}C_{2}\frac{\partial^4
\phi}{\partial x^{4}}-8n^{4}C_{2}(\frac{\partial
\phi}{\partial x}\frac{\partial^3
\phi}{\partial x^{3}}-\frac{1}{2}\frac{\partial^2
\phi}{\partial x^{2}}\frac{\partial^2
\phi}{\partial x^{2}})-C_{1}\frac{n^{4}}{4}\frac{\partial^{2}
\phi}{\partial {x^{2}}}\frac{\partial^{2}
\phi}{\partial {x^{2}}}\label{v13}\end{equation}

\begin{eqnarray}&&a_{02n}=-15n^{3}C_{2}\frac{\partial^4
\phi}{\partial x^{4}}+20n^{4}C_{2}\left(\frac{\partial
\phi}{\partial x}\frac{\partial^3
\phi}{\partial x^{3}}-\frac{1}{2}\frac{\partial^2
\phi}{\partial x^{2}}\frac{\partial^2
\phi}{\partial x^{2}}\right)+\frac{C_{1}}{4}\left(\frac{\partial^2
\phi}{\partial x^{2}}\frac{\partial^2
\phi}{\partial x^{2}}\right)\nonumber
\\ &&-8n^{5}C_{2}\int \left\lbrace\left(\frac{\partial
\phi}{\partial x}\right)^{2}\frac{\partial^3
\phi}{\partial x^{3}}-\frac{1}{2} \frac{\partial
\phi}{\partial x} \left(\frac{\partial^2
\phi}{\partial x^{2}}\right)^{2}\right\rbrace dx-C_{1}\frac{n^{5}}{4}\int\frac{\partial
\phi}{\partial {x}}\left(\frac{\partial^2
\phi}{\partial x^{2}}\right)^{2}dx\label{v14p}\end{eqnarray}

After a simple rearrangement in the last term
\begin{eqnarray}&&a_{02n}=-15n^{3}C_{2}\frac{\partial^4
\phi}{\partial x^{4}}+20C_{2}n^{4}\left(\frac{\partial
\phi}{\partial x}\frac{\partial^3
\phi}{\partial x^{3}}-\frac{1}{2}\frac{\partial^2
\phi}{\partial x^{2}}\frac{\partial^2
\phi}{\partial x^{2}}\right)\nonumber
\\ &&-8C_{2}n^{5}\left(\frac{\partial
\phi}{\partial x}\right)^{2}\frac{\partial^2
\phi}{\partial x^{2}}+\frac{C_{1}}{4}\left(\frac{\partial^2
\phi}{\partial x^{2}}\frac{\partial^2
\phi}{\partial x^{2}}\right)\label{v14}\end{eqnarray}

Therefore, with the help of eq.(\ref{v3}), eq.(\ref{4}) and eq.(\ref{5})
\begin{eqnarray}f_{4}=\sum_{n=1}^{\infty}\alpha\beta W_{4}\end{eqnarray}

where,
\begin{equation}\alpha=e^{\left(  -n\phi+\frac{n^{3}\Lambda^{2}}{24}\left( \frac{\partial
\phi}{\partial x}\right)^{2}-\frac{n^{2}\Lambda^{2}}{8}\left( \frac{\partial^{2}
\phi}{\partial x^{2}}\right)\right)}\end{equation}
\begin{eqnarray}\beta=e^{- 
\frac{np^{2}}{2•}\left( 1-\frac{n^{2}\Lambda^{2}}{12}\left( \frac{\partial^{2}
\phi}{\partial x^{2}}\right)\right)}\end{eqnarray}

To get back the Wigner's structure of the $\Lambda^{4}$ order solution, the exponential factors $W_{2}$ and $W_{4}$ should be expanded and terms upto $\Lambda^{4}$ order would be retained.
\begin{eqnarray}&&f_{4}=\sum_{n=1}^{\infty}e^{-n\left(  \phi+\frac{p^{2}}{2}\right) }\left(1+\frac{n^{3}\Lambda^{2}}{24}\left( \frac{\partial
\phi}{\partial x}\right)^{2}+ \frac{n^{6}\Lambda^{4}}{1152}\left( \frac{\partial
\phi}{\partial x}\right)^{4}+...\right)\nonumber\\&&\nonumber\\&& \left(1-\frac{n^{2}\Lambda^{2}}{8}\left( \frac{\partial^{2}
\phi}{\partial x^{2}}\right)+ \frac{n^{4}\Lambda^{4}}{128}\left( \frac{\partial^{2}
\phi}{\partial x^{2}}\right)^{2}+...\right)\nonumber\\&&\left(1+\frac{n^{3}\Lambda^{2}p^{2}}{24}\left( \frac{\partial^{2}
\phi}{\partial x^{2}}\right)+ \frac{n^{6}\Lambda^{4}p^{4}}{1152}\left( \frac{\partial^{2}
\phi}{\partial x^{2}}\right)^{2}+...\right)\nonumber\\&&\left(1+\Lambda^{4}\left(a_{02n}+a_{12n}\frac{np^{2}}{2}+a_{22n}{{\left(\frac{np^{2}}{2}\right)}^{2}}\right)+... \right) \end{eqnarray}
Collecting terms upto the order $\Lambda^{4}$

\begin{equation}f_{4}=\sum_{n=1}^{\infty}e^{-n(\phi+\frac{p^{2}}{2})}(C+D+E+F+G+H+I+J+K+L+M+N)\end{equation}
where
$$C=1+\Lambda^{2}\left( \frac{n^{3}}{24}\left( \frac{d\phi}{dx}\right)^{2}-\frac{n^{2}}{8}\frac{d^{2}\phi}{dx^{2}}+\frac{n^{3}p^{2}}{24}\frac{d^{2}\phi}{dx^{2}}\right) $$
$$D=-\frac{3n^{5}\Lambda^{4}}{320}\left( \frac{d\phi}{dx}\right) ^{2}\frac{d^{2}\phi}{dx^{2}}$$
$$E=\frac{5n^{4}\Lambda^{4}}{384}\left( \frac{d^{2}\phi}{dx^{2}}\right) ^{2}$$
$$F=\frac{n^{6}\Lambda^{4}}{1152}\left( \frac{d\phi}{dx}\right)^{4} $$
$$G=-\frac{n^{3}\Lambda^{4}}{128}\left( \frac{d^{4}\phi}{dx^{4}}\right)$$
$$H=\frac{n^{4}\Lambda^{4}}{96}\left( \frac{d^{3}\phi}{dx^{3}}\right)\left( \frac{d\phi}{dx}\right)$$
$$I=\frac{n^{6}\Lambda^{4}p^{2}}{576}\left( \frac{d\phi}{dx}\right) ^{2}\frac{d^{2}\phi}{dx^{2}}$$
$$J=-\frac{3n^{5}\Lambda^{4}p^{2}}{320}\left( \frac{d^{2}\phi}{dx^{2}}\right)^{2}$$
$$K=-\frac{n^{5}\Lambda^{4}p^{2}}{480}\left( \frac{d^{3}\phi}{dx^{3}}\right)\left( \frac{d\phi}{dx}\right)$$
$$L=\frac{n^{4}\Lambda^{4}p^{2}}{192}\left( \frac{d^{4}\phi}{dx^{4}}\right)$$
$$M=\frac{n^{6}\Lambda^{4}p^{4}}{1152}\left( \frac{d^{2}\phi}{dx^{2}}\right) ^{2}$$
$$N=-\frac{n^{5}\Lambda^{4}p^{4}}{1920}\left( \frac{d^{4}\phi}{dx^{4}}\right)$$

This process can be continued and  for the $\Lambda^{2j}$ order solution we get the following equation.
\begin{equation}
(p
\frac{\partial W_{2jn}}{\partial {x}}-\frac{\partial
\phi}{\partial {x}}\frac{\partial W_{2jn}}{\partial
{p}})f_{2j-2,n}\\
+\sum_{i=1}^{j}(-1)^{i+1}C_{i}\Lambda^{2i}\frac{\partial^{2i+1}
\phi}{\partial x^{2i+1}}
\frac{\partial^{2i+1} f_{2j-2,n}}{\partial p^{2i+1}}W_{2jn}=0\label{vv20}\end{equation}
where, $$f_{2jn}=\sum_{n=1}^{\infty}\prod_{k=1}^{j}W_{2kn}$$

The solution of the complete Wigner equation will then be obtained as
$$f_{n}=\sum_{n=1}^{\infty}\prod_{j=1}^{\infty}W_{2jn}$$

For j=1, we have already obtained $W_{2}$

For $j>1$,
$$W_{2jn}=\exp (U_{2jn})$$
where
$$U_{2jn}=\sum_{i=0}^{j}\Lambda^{2j} a_{ijn}({\frac{p\sqrt{n}}{\sqrt{2}}})^{2i}$$

Now we can collect all $\Lambda^{2j}$ order terms in eq.(\ref{vv20}) to obtain
\begin{equation}
(p
\frac{\partial W_{2jn}}{\partial {x}}-\frac{\partial
\phi}{\partial {x}}\frac{\partial W_{2jn}}{\partial
{p}})MR+S\\
+(-1)^{j+1}C_{j}\Lambda^{2j}\frac{\partial^{2j+1}
\phi}{\partial x^{2j+1}}
\frac{\partial^{2j+1}M}{\partial p^{2j+1}}RW_{2jn}=0\label{vvv20}\end{equation}
where $M=e^{-\frac{np^{2}}{2}}$, $R=f_{2j-2,n}/M$ and S is the $\Lambda^{2j}$ order contribution of the terms,
$$\sum_{i=1}^{j-1}(-1)^{i+1}C_{i}\Lambda^{2i}\frac{\partial^{2i+1}
\phi}{\partial x^{2i+1}}
\frac{\partial^{2i+1} f_{2j-2,n}}{\partial p^{2i+1}}W_{2jn}$$

For example, if we are interested in $\Lambda^{4}$($j=2$) order solution, only the first term of the series 
$$C_{1}\Lambda^{2}\frac{\partial^{3}
\phi}{\partial x^{3}}
\frac{\partial^{3} f_{2j-2,n}}{\partial p^{3}}W_{2jn}$$
will contribute to S.

$$S=-pC_{1}\Lambda^{4}\frac{n^{4}}{2•}\frac{\partial^{2}
\phi}{\partial x^{2}}\frac{\partial^{3}
\phi}{\partial x^{3}}
 f_{2,n}W_{2jn}+p^{3}C_{1}\Lambda^{4}\frac{n^{5}}{4•}\frac{\partial^{2}
\phi}{\partial x^{2}}\frac{\partial^{3}
\phi}{\partial x^{3}}
 f_{2,n}W_{2jn}$$
Let $q^{2}=np^{2}$

\begin{equation}
(\frac{q•}{\sqrt{n}•}
\frac{\partial W_{2jn}}{\partial {x}}-\sqrt{n}\frac{\partial
\phi}{\partial {x}}\frac{\partial W_{2jn}}{\partial
{q}})MR+S(x,q)\\
+n^{j+\frac{1•}{2•}}(-1)^{j+1}C_{j}\Lambda^{2j}\frac{\partial^{2j+1}
\phi}{\partial x^{2j+1}}
\frac{\partial^{2j+1}M}{\partial q^{2j+1}}RW_{2jn}=0\label{vvv20}\end{equation}
\begin{equation}p\frac{dW_{2jn}}{dx}=\frac{q•}{\sqrt{n}•}\sum_{0}^{j}\Lambda^{2j}(\frac{da_{0jn}}{dx}+...+\frac{da_{ijn}}{dx}\frac{q^{2i}}{2^{i}}+...)W_{2j}\label{v15}\end{equation}

\begin{equation}\frac{d\phi}{dx}\frac{dW_{2jn}}{dp}=\sqrt{n}\frac{d\phi}{dx}\sum_{0}^{j}\Lambda^{2j}(...+a_{ijn}(2i)\frac{q^{2i-1}}{2^{i}}+a_{i+1,j,n}(2i+2)\frac{q^{2i+1}}{2^{i+1}}+...)W_{2jn}\label{v15n}\end{equation}
Comparing the $\Lambda^{2j}$ order coefficient of $q^{2i+1}$ in  eq.(\ref{vvv20}), the following equation is obtained

\begin{equation}\frac{1}{2^{i}}\frac{da_{ijn}}{dx}=n\frac{i+1}{2^{i}}\frac{d\phi}{dx}a_{i+1,jn}+b_{ij}n^{j+1}+g_{ijn}\label{v15nn}\end{equation}

where the last term is the coefficient of  $q^{2i+1}$ of S and

\begin{equation}b_{ij}=-\frac{\partial^{2j+1} \phi}{\partial x^{2j+1}}D_{i}C_{j}\label{v16}\end{equation}
in which, $D_{i}=$ coefficient of $q^{2i+1}$ of the Hermite polynomial $He_{2j+1}(q)$

These equations are true for i=0 to i=j-1

For i=j,
\begin{equation}\frac{1}{2^{i}}\frac{\partial a_{ijn}}{\partial x}=-n^{j+1}\frac{\partial^{2j+1} \phi}{\partial x^{2j+1}}C_{j}D_{j}\label{v17}\end{equation}

This is a first order equation and can be easily solved to obtain the value of $a_{jj}$. Using this result, all the equations of this group can be successively solved to find out the remaining coefficients. Finally, $a_{ijn}$ can be expressed in the following compact form. 
\begin{equation}a_{ijn}=\sum_{k=0}^{j-i}c_{ikj}n^{j+k+1}\label{v17}\end{equation}
$c_{ikj}$ can be identified from the expression of $a_{ijn}$. For example, the coefficents of $n^{3}$,$n^{4}$ and $n^{5}$ of $a_{02n}$ in eq.(\ref{v14}) can be identified as $c_{002}$,$c_{012}$ and $c_{022}$ respectively.

Therefore, the distribution corrected upto $\Lambda^{2j\prime•}$ order is 
$$exp[\sum_{j=1}^{j\prime}\sum_{i=0}^{j}\sum_{k=0}^{j-i}\Lambda^{2j}\frac{p^{2i}}{2^{i}•}c_{ikj}\partial^{i+k+j+1}]f$$
$$f=\sum_{n=1}exp[-n(\frac{p^{2}}{2•}+\phi)]=\frac{1•}{exp[(\frac{p^{2}}{2•}+\phi)]-1•}$$
($\partial$ = derivative with respect to $\phi$)
\section{Conclusion}

In this work the all order solution of the Wigner equation for the bosons has been derived in the presence of mean field. In the original work of Wigner he calculated the second order quantum correction to the Gibbs-Boltzmann distribution and applied it to  calculate the thermodynamic properties of the many-body system. It is evident that as we increase the density and decrease the temperature of the system we approach the situation where quantum effect is more important and to describe it properly we need to add more higher order terms to the Wigner equation. In this condition the indistinguishibility factor of the particles should be incorporated for the completeness of the problem and we have taken care of that factor in our calculation. 

For example, at finite temperature, the number density, pressure exerted by bosons can be easily calculated by taking the appropriate moments of the quantum corrected Bose-Einstein distribution function derived in this article. 

$$n_{cb}=\int exp[\sum_{j=1}^{j\prime}\sum_{i=0}^{j}\sum_{k=0}^{j-i}\Lambda^{2j}\frac{p^{2i}}{2^{i}•}c_{ikj}\partial^{i+k+j+1}]fdp$$
where $n_{cb}$ is the quantum corrected boson density and 
$$f=\sum_{n=1}exp[-n(\frac{p^{2}}{2•}+\phi)]=\frac{1•}{exp[(\frac{p^{2}}{2•}+\phi)]-1•}$$

We can calculate the expression upto the $\Lambda^{4}$ order correction.
$$n_{cb}=\int exp[\Lambda^{2}(c_{001}\partial^{2}+c_{011}\partial^{3}+\frac{p^{2}}{2}c_{101}\partial^{3})+\Lambda^{4}(c_{002}\partial^{3}+c_{012}\partial^{4}+c_{022}\partial^{5}+\frac{p^{2}}{2}(c_{102}\partial^{4}+c_{112}\partial^{5})+\frac{p^{4}}{4}c_{202}\partial^{5})]fdp$$
The exponential will be expanded and terms upto the $\Lambda^{4}$ order will be retained

$$n_{cb}=\int [1+\Lambda^{2}(c_{001}\partial^{2}+c_{011}\partial^{3}+\frac{p^{2}}{2}c_{101}\partial^{3})+$$$$\frac{\Lambda^{4}}{2}(c_{001}\partial^{2}+c_{011}\partial^{3}+\frac{p^{2}}{2}c_{101}\partial^{3})^{2}+\Lambda^{4}(c_{002}\partial^{3}+c_{012}\partial^{4}+c_{022}\partial^{5}+\frac{p^{2}}{2}(c_{102}\partial^{4}+c_{112}\partial^{5})+\frac{p^{4}}{4}c_{202}\partial^{5})]fdp$$
Performing the integration, we obtain

$$n_{cb}=n_{b}+\Lambda^{2}(c_{001}\frac{\partial^{2}n_{b}}{\partial \phi^{2}}+c_{011}\frac{\partial^{3}n_{b}}{\partial \phi^{3}}+2^{1/2}c_{101}\Gamma(3/2)\frac{\partial^{3}}{\partial \phi^{3}•} g_{(3/2)}(e^{-\phi}))+$$$$\Lambda^{4}(\frac{c^{2}_{001}}{2}\frac{\partial^{4}n_{b}}{\partial \phi^{4}}+\frac{c^{2}_{011}}{2}\frac{\partial^{6}n_{b}}{\partial \phi^{6}}+c_{000}c_{011}\frac{\partial^{5}n_{b}}{\partial \phi^{5}}+2^{1/2}c_{000}c_{101}\Gamma(3/2)\frac{\partial^{5}}{\partial \phi^{5}•} g_{(3/2)}(e^{-\phi})+$$$$+2^{1/2}c_{011}c_{101}\Gamma(3/2)\frac{\partial^{6}}{\partial \phi^{6}•} g_{(3/2)}(e^{-\phi})+2^{1/2}\frac{c^{2}_{101}}{2}\Gamma(5/2)\frac{\partial^{6}}{\partial \phi^{6}•} g_{(5/2)}(e^{-\phi}))+$$$$\Lambda^{4}(c_{002}\frac{\partial^{3}n_{b}}{\partial \phi^{3}}+c_{012}\frac{\partial^{4}n_{b}}{\partial \phi^{4}}+c_{022}\frac{\partial^{5}n_{b}}{\partial \phi^{5}}+2^{1/2}c_{102}\Gamma(3/2)\frac{\partial^{4}}{\partial \phi^{4}•} g_{(3/2)}(e^{-\phi})+2^{1/2}c_{112}\Gamma(3/2)\frac{\partial^{5}}{\partial \phi^{5}•} g_{(3/2)}(e^{-\phi})+$$$$2^{1/2}c_{202}\Gamma(5/2)\frac{\partial^{5}}{\partial \phi^{5}•} g_{(5/2)}(e^{-\phi})) $$


We have used the following integral to obtain the above expression

$$\int_{-\infty}^{\infty•}\frac{p^{2k}dp•}{exp(\frac{p^{2}}{2•})z^{-1}-1•}=2^{k+1/2}\Gamma(k+1/2)g_{(k+1/2)}(z)$$
Therefore, we have been able to calculate the quantum corrected density function upto the $\Lambda^{4}$ order. If the quantum corrections are ignored, we get back the usual density function of the bosons. 

The process is simple and can be continued for any order. Similarly, we can also calculate other physical quantities like pressure by taking the proper moments of the distribution function. It is observed that the higher order terms of the Wigner equation contain higher derivatives of the potential function. Therefore, the potential should be smooth enough so that the convergence of the problem is achieved.

Finally, it can be concluded that this article extends the phase space formulation of Wigner to the system of bosons with all higher order corrections and may be applied to probe the properties of the bosons and bose condensates at finite temperature in the presence of external potential.


\newpage
\begin{thebibliography}{99}
\bibitem{kn:neu} Von Neumann, Gott. Nachr.{\bf{273}} (1927).
\bibitem{kn:blo} Bloch, Zeits. f. Physik. {\bf{74}}, 295 (1932).
\bibitem{kn:wigner} E. Wigner, Phys. Rev. {\bf{40}}, 749 (1932).
\bibitem{kn:kirk} J. G. Kirkwood, Phys. Rev. {\bf{44}}, 31 (1931).
\bibitem{kn:uhl} G. E. Uhlenbeck and E. Beth, Physica {\bf{3}}, 729 (1936) j {\bf{4}}, 915
(1937).
\bibitem{kn:deb} J. de Boer, Amsterdam Dissertations (1940).
\bibitem{kn:green} H.S. Green, J. Chem. Phys. {\bf{19}}, 955 (1951).
\bibitem{kn:moy}J.E. Moyal,  Proc. Cambridge Phil. Soc. {\bf{45}}, 99
(1949)
\bibitem{kn:bon}M. Bonitz, Quantum Kinetic Theory (Teubner, Stuttgart, 1998)
\bibitem{kn:tata}V.I. Tatarskii, Sov. Phys. Usp. {\bf{26}}, 311
(1983)
\bibitem{kn:carr}P. Carruthers, F. Zachariasen, Rev.
Mod. Phys. {\bf{55}}, 245 (1983)
\bibitem{kn:hill}M. Hillery, R.F. O’Connell, M.O. Scully, E.P. Wigner, Phys. Rep. {\bf{106}}, 121 (1984)
\bibitem{kn:lee}H.W. Lee, Phys.
Rep., {\bf{259}},147 (1995)
\bibitem{kn:saso}V. N. Sasonov and A. A. Stuchebrukhov,Chem. Phys.        {\bf{56}}, 391 (1981).
\bibitem{kn:kors}H. J. Korsch and M. V. Berry, Physica {\bf{3D}}, 627 (1981).
\bibitem{kn:klim} Yu. L. Klimontovich, Dokl. Akad. Nauk SSSR {\bf{108}}, 1033
(1956) .
\bibitem{kn:soto}F. Soto and P. Claverie, Physica, {\bf{109A}}, 193 (1981).
\bibitem{kn:sch}W.P.Schleich, Quantum Optics in Phase Space, WILEY-VCH, (2001).
\bibitem{kn:zoppi} Barocchi, M. Neumann, and M. Zoppi, Phys. Rev. A {\bf{31}}, 4015 (1985).
\bibitem{kn:singh}A. K. Singh and S. K. Sinha, Phys. Rev. A {\bf{30}}, 1078 (1984).
\bibitem{kn:koro}V. Korobkin and V. N. Sazonov, Sov. Phys. JETP {\bf{54}}, 636 (1981).
\bibitem{kn:hansen}J. P. Hansen and P. Vieillefosse, Phys. Lett. A \textbf{53}, 187 (1975).
\bibitem{kn:jan1}A. Alastuey and B. Jancovici, Phys. A (Amsterdam) \textbf{97}, 349 (1979).
\bibitem{kn:bas}M.J. Bastiaans,  Opt. Commun. \textbf{25}, 26 (1978).
\bibitem{kn:uhl1} G. E. Uhlenbeck and L. Gropper, Phys. Rev. {\bf{41}}, 79 (1932).
\bibitem{kn:schram}K.Schram and B.R.A. Nijboer, Physica,{\bf{25}}, 733 (1959).
\bibitem{kn:jan4} A. Alastuey and B. Jancovici, Physica A,  \textbf{97A}, 349(1979).
\bibitem{kn:jan3} A. Alastuey and B. Jancovici, Physica A,  \textbf{102A}, 327(1980).
 

\bibitem{kn:dahl}J.P. Dahl  Can. J. Chem. {\bf{87}}, 784(2009)
\bibitem{kn:blan} P.Blanchard, J.M.Gracia-Bonda and J.C.Varilly, Int. J. Quantum Chem. {\bf{112}}, 1134(2012)
\bibitem{kn:smer} A.Smerzi, Phys. Rev. A, {\bf{52}}, 4365 (1995)
\bibitem{kn:brand}Brandon P van ZylJ, Phys. A: Math. Theor, {\bf{45}},  315302 (2012).
\bibitem{kn:sell}
J.M.Sellier,M. Nedjalkov,I. Dimov, Physiscs Reports, {\bf{577}}, 1 (2015)
\bibitem{kn:anirban} A.Bose and M.S. Janaki, Phys. Plasmas, {\bf{19}}, 072101 (2012).

 

%
%




\end{thebibliography}
\end{document}